\documentclass[%
superscriptaddress,
array,
amsmath,amssymb,
aps,
longbibliography,
]{revtex4-1}

\renewcommand{\toprule}{\hline}
\newcommand{\midrule}{\hline}
\newcommand{\bottomrule}{\hline}
\usepackage[switch*, modulo]{lineno}

\usepackage[colorinlistoftodos]{todonotes}

\usepackage{enumitem}

\usepackage{datatool-base}

\usepackage{titlesec}
\titlespacing{\section}{0pc}{1pc}{0.5pc}
\titlespacing{\subsection}{0pc}{1pc}{0.5pc}

\usepackage{hyperref}

\begin{document}

\title{The ILD Detector: A Versatile Detector for an Electron-Positron Collider at Energies up to 1~TeV 
}



%
%
\author{H.~Abramowicz} \affiliation{Department of Particle Physics, School of Physics and Astronomy, Tel Aviv University, Israel}
\author{D.~Ahmadi} \affiliation{Vrije Universiteit Brussel, Brussel, Belgium}
\author{J.~Alcaraz} \affiliation{Centro de Investigaciones Energeticas, Medioambientales y Tecnologicas, Madrid, Spain}
\author{O.~Alonso} \affiliation{Electronic and Biomedical Engineering Department, University of Barcelona, Spain}
\author{L.~Andricek} \affiliation{MPG Halbleiterlabor, Garching, Germany}
\author{J.~Anguiano} \affiliation{Department of Physics and Astronomy, University of Kansas, USA}
\author{O.~Arquero} \affiliation{Centro de Investigaciones Energeticas, Medioambientales y Tecnologicas, Madrid, Spain}
\author{F.~Arteche} \affiliation{Instituto Tecnológico de Aragón, Zaragoza, Spain}
\author{D.~Attie} \affiliation{CEA/Irfu Universit\'e Paris Saclay, France}
\author{O.~Bach} \affiliation{Deutsches Elektronen-Synchrotron, Hamburg, Germany}
% ...


\begin{abstract}
The International Large Detector, ILD, is a detector concept for an experiment at a future high energy lepton collider. The detector has been optimised for precision physics in a range of energies from 90~GeV to about 1~TeV. ILD features a high precision, large volume combined silicon and gaseous tracking system, together with a high granularity calorimeter, all inside a central solenoidal magnetic field. The paradigm of particle flow has been the guiding principle of the design of ILD. ILD is based mostly on technologies which have been demonstrated by extensive research and test programs. The ILD concept is proposed both for linear and circular lepton collider, be it at CERN or elsewhere. The concept has been developed by a group of nearly 60 institutes from around the world, and offers a well developed and powerful environment for science and technology studies at lepton colliders. In this document, the required performance of the detector, the proposed implementation and the readiness of the different technologies needed for the implementation are discussed.

\end{abstract}
\



\date{\today}




\maketitle


\setcounter{page}{1}



\section{\label{sec:level1}Introduction}
The International Large Detector, ILD, is a proposal for a detector at a future electron-positron collider, for energies up to about 1~TeV. In this paper, the considerations which have guided the ILD concept group in the design of the detector are summarised. The main 
challenges for the realisation of the concept are described, and possible technological solutions are sketched. The ILD concept is developed by a broad and international community of scientists. 

 It is the intention of the ILD group to develop integrated detector concepts for future high-energy electron-positron colliders, both linear and circular colliders, such as ILC~\cite{Behnke:2013xla}, LCF~\cite{abramowicz_linear_2025}, and FCC-ee~\cite{Andre:2025bpv}. ILD is also studying how its concept would need to evolve to be usable at new proposals like the asymmetric collider proposal HALHF~\cite{HALHF:2025hqj}. ILD can be realised at any of the sites proposed for any of the different 
 collider proposals, be it at CERN or elsewhere.
 
 ILD was originally conceived as part of the ILC project. The ambitious requirements of the ILC detectors sparked a world-wide {{R\&D}} program to develop and demonstrate the different technologies needed~\cite{RDliaision}. The scope and the needs of the R\&D required for detectors at a future collider have been more recently summarised in a report to ECFA \cite{Contardo:2023qzs}.
Following the 2019 update of the European Strategy, a new generation of R\&D groups have been initiated, the so-called DRD collaborations. The former R\&D collaborations for the most part have been integrated into the DRD groups. The DRD groups are hosted by CERN, but are managed individually. 
 The ILD concept group from its beginning has collaborated very closely with these {{R\&D}} groups, and has organised the needed {{R\&D}} work through and with the {{R\&D}} collaborations. This close cooperation will continue also with the new mode of operation.

\section{The ILD Detector Design: Requirements}
ILD has been conceived and designed to study Higgs physics in great detail, and to improve our general understanding of electroweak and top-quark physics at high energies. The state-of-the-art regarding physics at such an high energy lepton collider has been summarized in the recent ECFA study report
\cite{ECFA-studyreport}. The science case is strongly dominated by the quest for high precision in measurements of the properties of 
the Higgs boson, the weak gauge bosons, and the top quark (see for example~\cite{ECFA-studyreport},\cite{Fujii:2017vwa},\cite{ILCESU1}, \cite{InternationalLinearColliderInternationalDevelopmentTeam:2021guz} ). 

The ultimate goal of 
this experimental program
is a much improved understanding of electroweak physics, and the interpretation of this in view of new insights into possible extensions of the Standard Model of particle physics.

The design of the ILD is driven by this quest for precision measurements, and its realisation through the concept of ``particle flow''. In this approach, all particles in an event, both charged and neutral, are individually reconstructed as much as possible. This requires a detector which can distinguish single particles even within dense jets, and which can find and separately measure neutral and charged particles. For the detector, this implies that particle separation power is receiving a lot of attention, and that in particular the calorimeters are designed with very high granularity, both in the transverse and 
longitudinal directions, and that the linking between the tracking and the calorimeter systems should be as efficient as possible. Particle flow is particularly important at higher center-of-mass energies, where events are in general less spherical and more jet-like. The design of the ILD detector is described in detail in \cite{ILDConceptGroup:2020sfq}.


The design drivers of the ILD detector can be summarized by the following requirements: 
\begin{itemize}[noitemsep,topsep=2pt]
    \item {\bf Impact parameter resolution:}  An impact parameter resolution of $ 5~\mu \mathrm{m} \oplus 10~\mu \mathrm{m} / [ p~({\mathrm{GeV}/c})\sin^{3/2}\theta$] has been defined as a goal, where $\theta$ is the angle between the particle and the beamline. 
    This ensures excellent identification of displaced vertices for the identification of e.g. b- and c-quarks.
    \item {\bf Momentum resolution:} An inverse momentum resolution of $\Delta (1 / p) = 2 \times 10^{-5}~\mathrm{(GeV/c)^{-1}}$ asymptotically at high momenta should be reached with the combined silicon - TPC tracker. Maintaining excellent tracking efficiency and very good momentum resolution at lower momenta will be achieved by an aggressive design to minimise the detector's material budget.
    This requirement is driven by the Higgs recoil mass measurement, ensuring that it's resolution is not dominated by the tracking performance.
    \item {\bf Jet energy resolution:} Using the paradigm of particle flow a jet energy resolution $\Delta E/ E = 3\%$ or better for light flavour jets should be reached. The resolution is defined in reference to light-quark jets, as the R.M.S. of the inner $90\%$ of the energy distribution. This resolution allows hadronic decays of the W, Z and Higgs bosons to be distinguished on a statistical basis.
    \item{ 
    {\bf Hermiticity:} The detector should cover as much as possible of the solid angle around the collision point, including in the very forward region. Ideally, the only uncovered regions should be the in- and out-going beampipes. This allows efficient identification of very forward-going particles, which, if missed, would restrict the ability to probe missing energy signals.
    }
    \item {\bf Readout:} The detector readout should avoid using a hardware trigger, ensuring maximal efficiency for all possible event topologies.
\end{itemize}

\begin{figure}[tb]
 \begin{center}
 \begin{tabular}{lr}
 \includegraphics[width=0.48\hsize,clip]{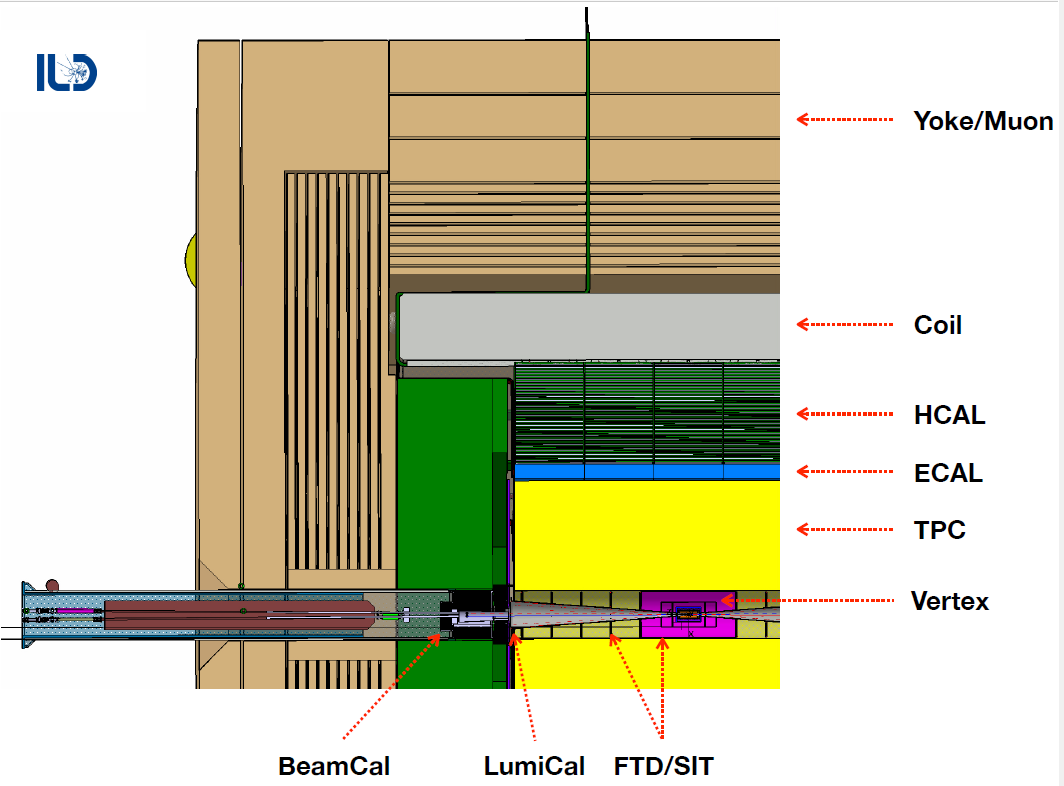} & 
 \includegraphics[width=0.35\hsize]{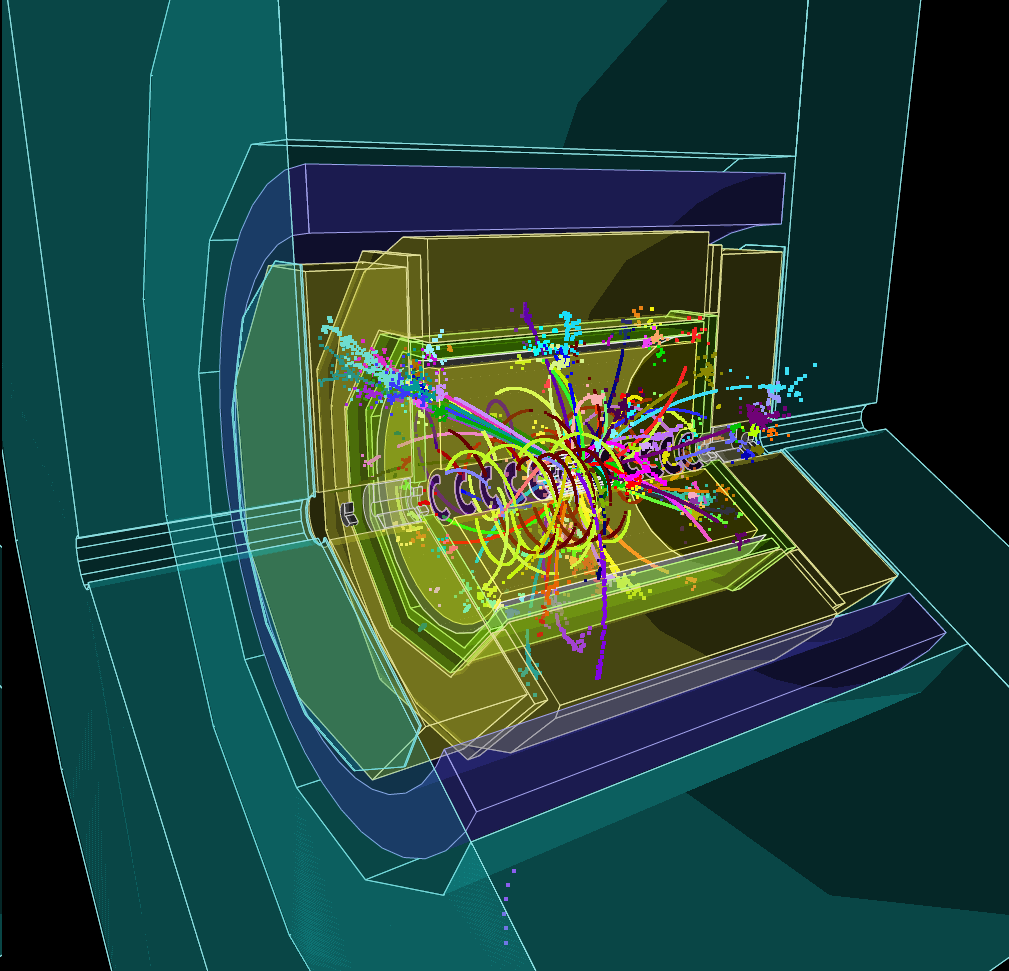}
 \\
 \end{tabular}
\caption{Left: Single quadrant view of the ILD detector. Right: Event display of a simulated hadronic decay of a $t \bar t$ event in ILD. The colouring of the tracks show the results of the reconstruction, each colour corresponding to a reconstructed particle.
\label{fig:ILD}}
 \end{center}
 \end{figure}

\subsection{Specifics for ILD at a Linear Collider}

ILD was originally conceived for use at a linear collider, in particular, the ILC ~\cite{Behnke:2013xla}. The ILC will operate in a so-called bunch-train mode, in which bunches spaced by a few 100~ns are combined into trains of 1 or 2~k bunches, 
which repeat at 5 or 10~Hz. The ILC anticipates a maximum center-of-mass energy of around 1~TeV. These boundary conditions have a profound impact on the design of the detector: 
\begin{itemize}[noitemsep,topsep=2pt]
    \item The relatively long inter-bunch time make it easy to distinguish between bunches and to uniquely assign objects in the detector to a bunch crossing. 
    \item The long time between bunch trains allows a thermal management, which, in most cases, can operate with a minimum of active cooling. Large parts of the detector can be operated in the so-called power-pulsing mode, in which power-hungry components are only activated during bunch trains. 
    \item The long intervals between bunch trains also allow a local buffering of data during bunch trains, and a readout during the inter-bunch intervals. This opens the way to a triggerless operation of the detector. 
    \item The very high final energy of 1~TeV implies a relatively thick iron calorimeter and iron return yoke, to be able to efficiently reconstruct events up to highest energies. 
    \item The very high focusing of the beams at a linear collider result in intense beam-strahlung at the IP, whose effect on inner detector systems must be considered in their design.
    \item The beam optics of linear colliders allow a relatively large distance between the IP and the first beam elements. The central part of the detector thus can be designed without heavy objects, for example, magnets, intruding into the central detector region. 
\end{itemize}

Studies are ongoing into how the ILD design would change for use at an asymmetric collider, such as the HALHF proposal. 
First results indicate that this would not significantly change the performance of the detector. 

\subsection{Specifics for ILD at a Circular Collider}

ILD proposes to use a derivative of the detector at the FCC-ee collider proposal. The FCC-ee does not operate in a bunch train mode, but in a continuous mode. The time interval between collisions is as short as 
20~ns, significantly less than at the ILC. 
\begin{itemize}[noitemsep,topsep=0pt]
    \item The collision repetition rate at the FCC-ee forbids the use of power pulsing, signficantly increasing (by about a factor of 100) the power dissipation in detectors, compared to linear colliders.
    \item The continuous operation of the collider makes a triggerless operation significantly more challenging. 
    \item The lower top-energy of the colliders allows for a design of a more compact (and thus somewhat cheaper) detector.
    \item The final beamline elements are significantly closer to the IP, and will intrude into the active detector region. A very careful design of the innermost part of the detector is needed to minimize the backgrounds in the detector, and to maximise the detector acceptance. 
    \item As the beams are less tightly focused, the beamstrahlung per collision is less intense. Due to the larger number of collisions per time however, the integral background levels are comparable or larger that those at a linear collider. 
    \item The FCC-ee proposal anticipates 
    significant running at Z energies. This puts very different requirements on the detector in particular in the innermost region. Operation of ILD at the very high rates expected at the Z will need significantly more study.  Z operation might also imply a weaker magnetic field. 
\end{itemize}

\section{The ILD Detector System}

The ILD detector is a multi-purpose detector in which the different requirements are addressed by a combination of different sub-detector systems. The optimization for ultimate precision in the reconstruction of charged and neutral particles requires that all major systems are contained within a strong solenoidal magnetic field, of 3.5\,T, 
separates the impact into the calorimeter of nearby charged and neutral particles, 
and removes low-energy background from the main part of the detector. Ultimate precision also requires that as little material as possible is introduced 
before the calorimeters,
which implies that the tracking and calorimeter systems should be within the coil.
Over the past few years, an intense effort has been undertaken by the ILD concept group,
to optimize the size of the ILD detector~(see \cite{ILDConceptGroup:2020sfq} for a review of the optimization).

A quadrant view of the detector model is shown in Fig.~\ref{fig:ILD} (left), together with an event display in this detector of a $t \bar t$ event (right).

The ILD concept from its inception has been open to new technologies. 
No final decision on subdetector technologies has been taken at this time, and in many cases several options are currently under consideration. For any technology to be accepted by ILD however its capabilities have to have been demonstrated experimentally, including demonstartion of its performance with prototypes and under as realistic conditions as possible.


The main parameters of the ILD detector are summarised in Table~\ref{ild:tab:barrelpara}, together with the different technological options currently under consideration. 

\begin{table}[th]
    \centering
    \begin{tabular}{|l|l|c|c|p{4cm}|}
    \toprule
        {\bf Technology} & {\bf Detector} & {\bf Start (mm)}   & {\bf Stop (mm)} & {\bf Comment} \\
        \midrule
Pixel detectors & Vertex & $r_{in}=16$   & $r_{out}=58$   & 3 double layers of silicon pixels \\
& Forward tracking  & $z_{in}=220$ & $z_{out}=371$ & 2 Pixel disks \\
 & SIT    & $r_{in}=153$  & $r_{out}=303$  & 2 double layers of Si pixels            \\
\midrule
Silicon strip & Forward tracking  & $z_{in}=645$ & $z_{out}=2212$ & 5 layers of Si strips\\
                & SET    & $r_{in}=1773$ & $r_{out}=1776$ & 1 double layer of Si strips           \\
                & & & & \\
\midrule
Gaseous tracking & TPC & $r_{in}=329$ & $r_{out}=1770$ & MPGD readout, 220 points along the track, Alternative: pixel readout \\
\midrule
Silicon tungsten calorimeter & ECAL option& $r_{in}=1805$ & $r_{out}=2028$ & 30 layers of $5\times 5~\mathrm{mm}^2$ pixels \\
& ECAL EC option & $z_{in}=2411$ & $z_{out}=2635$ & 30 layers of $5\times 5~\mathrm{mm}^2$ pixels \\
& Luminosity calorimeter &$r_{in}=83$ & $r_{out}=194$& 30 layers\\
& &$z_{in}=2412$ & $z_{out}=2541$& \\
\midrule
Diamond tungsten or 
& Beam calorimeter &$r_{in}=18$ &$r_{out}=140$& 30 layers\\
GaAs calorimeter && $z_{in}=3115$&$z_{out}=3315$&\\
\midrule
SiPM-on-Tile & ECAL alternative   & $r_{in}=1805$ & $r_{out}=2028$ & 30 layers, 5~mm strips, crossed\\
& ECAL EC alternative& $z_{in}=2411$ & $z_{out}=2635$ & 30 layers, 5~mm strips, crossed\\
             & HCAL option   & $r_{in}=2058$ & $r_{out}=3345$ & 48 layers, $3\times 3~\mathrm{cm}^2$ pixels\\
             & HCAL EC option& $z_{in}=2650$ & $z_{out}=3937$ & 48 layers, $3\times 3~\mathrm{cm}^2$ pixels\\
\midrule
RPC          & HCAL option   & $r_{in}=2058$ & $r_{out}=3234$ & 48 layers, $1 \times 1 ~\mathrm{cm}^2$ pixels \\
& HCAL EC option & $z_{in}=2650$ & $z_{out}=3937$ & 48 layers, $1 \times 1~\mathrm{cm}^2$ pixels\\
\midrule
SiPM on scintillator bar & Muon & $r_{in}=4450$ & $r_{out}=7755$ & 14 layers \\
& Muon EC & $z_{in}=4072$ & $z_{out}=6712$ & up to 12 layers \\

\bottomrule
    \end{tabular}
    \caption{Key parameters of the ILD detector. All numbers from~\cite{Behnke:2013lya}. ``Start'' and ``Stop'' refer to the minimum and maximum extent of subdetectors in radius and/or $z$-value.}
    \label{ild:tab:barrelpara}
\end{table}

\subsection{Vertexing System}
The system closest to the interaction region is a pixel detector designed to reconstruct decay vertices of short lived particles with great precision. ILD has chosen a system consisting of three double layers of pixel detectors. The innermost layer is only half as long as the others to reduce the exposure to background hits. Each double  layer will provide a spatial resolution around 3~$\mu\mathrm{m}$ at a pitch of about 22~$\mu\mathrm{m}$, and a timing resolution per double layer of around 2--4~$\mu\mathrm{s}$ (see e.g. \cite{Winter:2012ms, Llopart_2022}). To this end, the two layers in one double layer are optimized individually, one towards best spatial resolution, the other towards excellent timing resolution. R\&D is directed towards improving this even further, to a point which would allow hits from individual bunch crossings to be resolved.

Over the last 10 years the MAPS technology has matured close to a point where all the requirements (material budget, readout speed, granularity) needed for an ILC detector can be met. The technology has seen a first large scale use in the STAR vertex detector~\cite{ild:bib:VTXcps3}, and more recently in the upgrade of the ALICE vertex detector. 
MAPS technology in general is undergoing very rapid progress and development, with many promising avenues being explored. To minimize the material in the system, sensors are routinely thinned to 50~$\mu{\mathrm m}$. 

Advancements in technology make it possible to revisit the baseline geometry (three double-sided layers). In particular, stitched sensors, explored within the ALICE-ITS3 upgrade, offer the potential to incorporate bent sensors to optimize the material budget. However, significant challenges remain in demonstrating the feasibility of this approach in the context of a lepton collider, including azimuthal acceptance and bending at a radius close to 12~mm.

Other technologies under consideration for ILD are DEPFET, which is also currently being deployed in the Belle II vertex detector ~\cite{Luetticke:2017zpx}, fine pitch CCDs ~\cite{fineCCD}, and also less mature technologies such as SOI (Silicon-on-insulator) and Chronopix~\cite{RDliaision}.
Very light weight support structures have been developed, which bring the goal of 0.15\% of a radiation length per layer within reach~\cite{PLUME:2011rwc}. Such structures are now used in the Belle II vertex detector.

\begin{figure}[t]
    \centering
    \includegraphics[width=0.35\hsize]{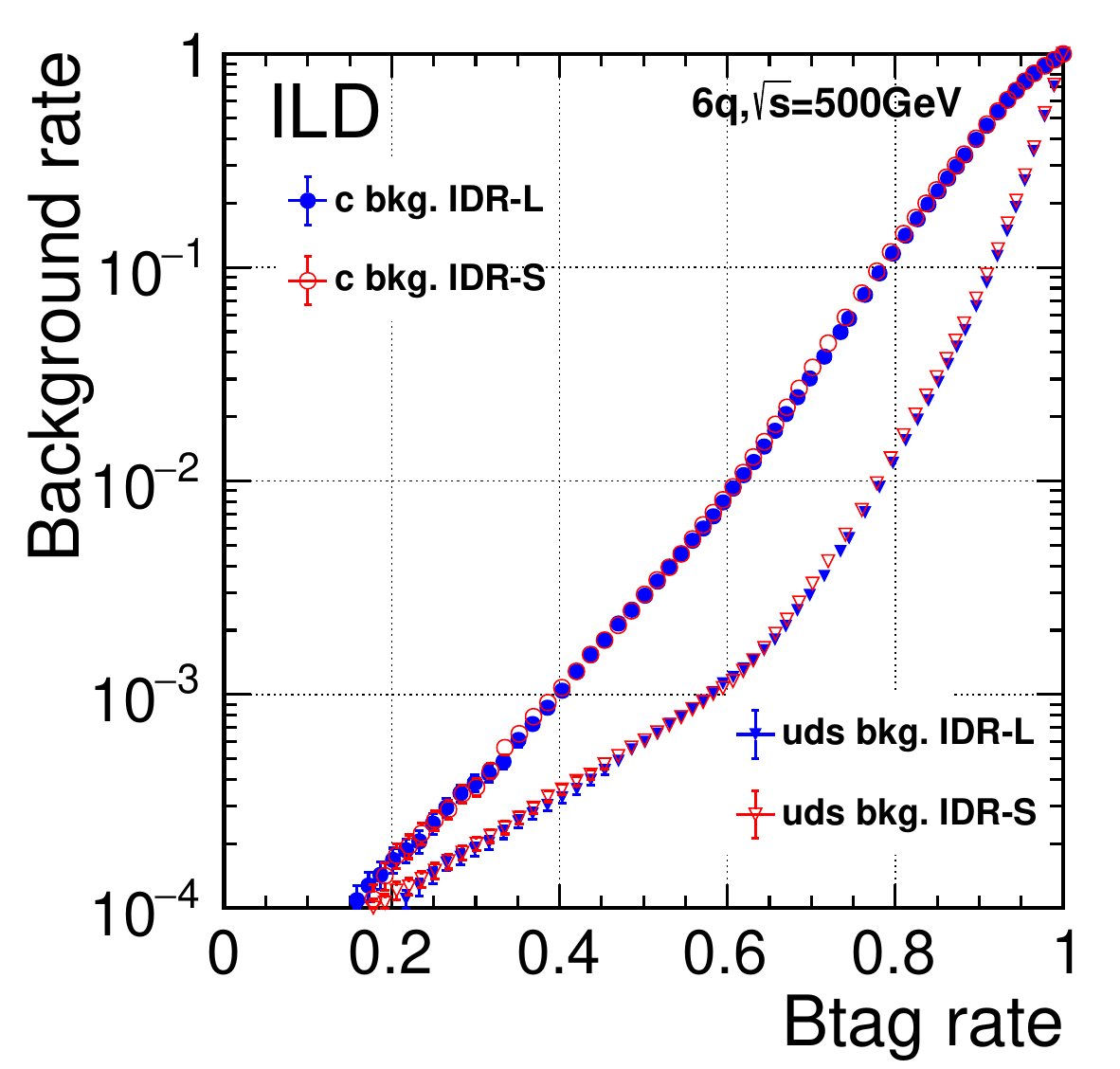}
    \includegraphics[width=0.35\hsize]{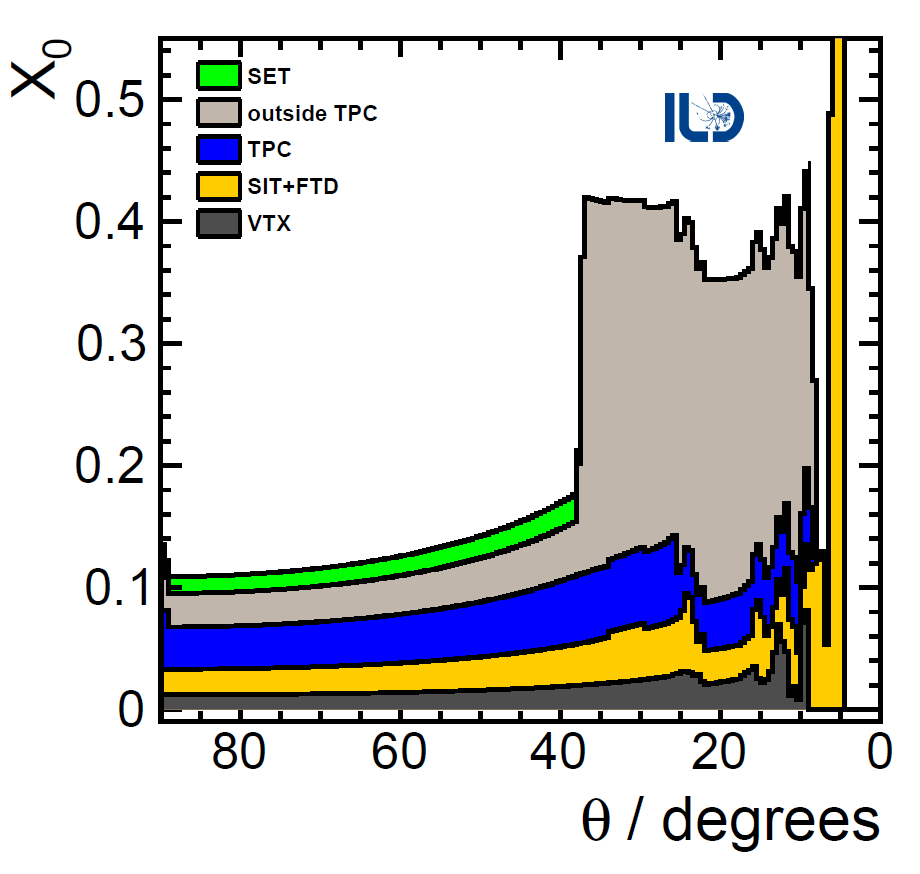}
    \caption{Left: Purity of the flavour tag as a function of the efficiency, for different flavours tagged. Right: Cumulative material budget in ILD up to the calorimeter, in fraction of a radiation length. Figures are taken from \cite{ILDConceptGroup:2020sfq}.}
    \label{fig-btag}
\end{figure}

In Fig.~\ref{fig-btag} the purity of the flavour identification in ILD is shown as a function of its efficiency.
The performance for b-jet identification is excellent, and charm-jet identification is also good, providing a purity of about 70\% at an efficiency of 60\%.
 The system also allows the accurate determination of the charge of displaced vertices, and contributes strongly to the low-momentum tracking capabilities of the overall system, down to a few 10s of\,MeV. An important aspect of the system leading to superb flavour tagging is the small amount of material in the tracker. This is shown in Fig.~\ref{fig-btag} (right).

\subsection{Tracking System}

ILD has decided to approach the problem of charged particle tracking with a hybrid solution, which combines a high resolution time-projection chamber (TPC) with a few layers of strategically placed strip or pixel detectors before and after the TPC (for a recent review, see \cite{Kaminski:2023vfc}).
The TPC 
will fill a large volume about 4.6\,m in length, spanning radii from 33 to 180\,cm. In this volume the TPC provides up to 220 three dimensional points for continuous tracking with a single-hit resolution of better than 100~$\mu\mathrm{m}$ in $r \phi$, and about 1\,mm in $z$. This high number of points allows a reconstruction of the charged particle component of the event with high accuracy, including the reconstruction of secondaries, long lived particles, kinks, etc. For momenta above 100\,MeV, and within the acceptance of the TPC, greater than 99.9\% tracking efficiency has been found in events simulated realistically with full backgrounds. At the same time the complete TPC system will introduce only about 10\% of a radiation length into the detector~\cite{Diener:2012mc}. 

\begin{figure}
    \centering
    \begin{tabular}{cc}

    \includegraphics[width=.41\hsize]{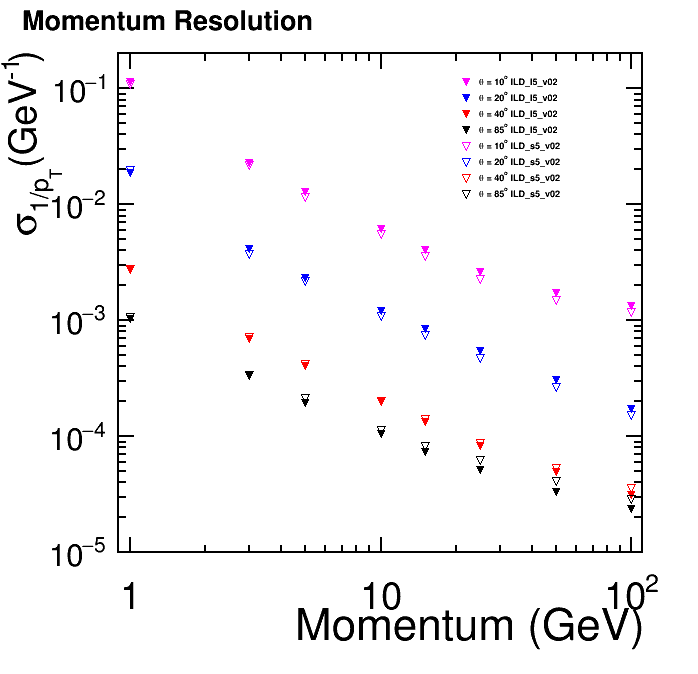} &
        \includegraphics[width=.47\hsize]{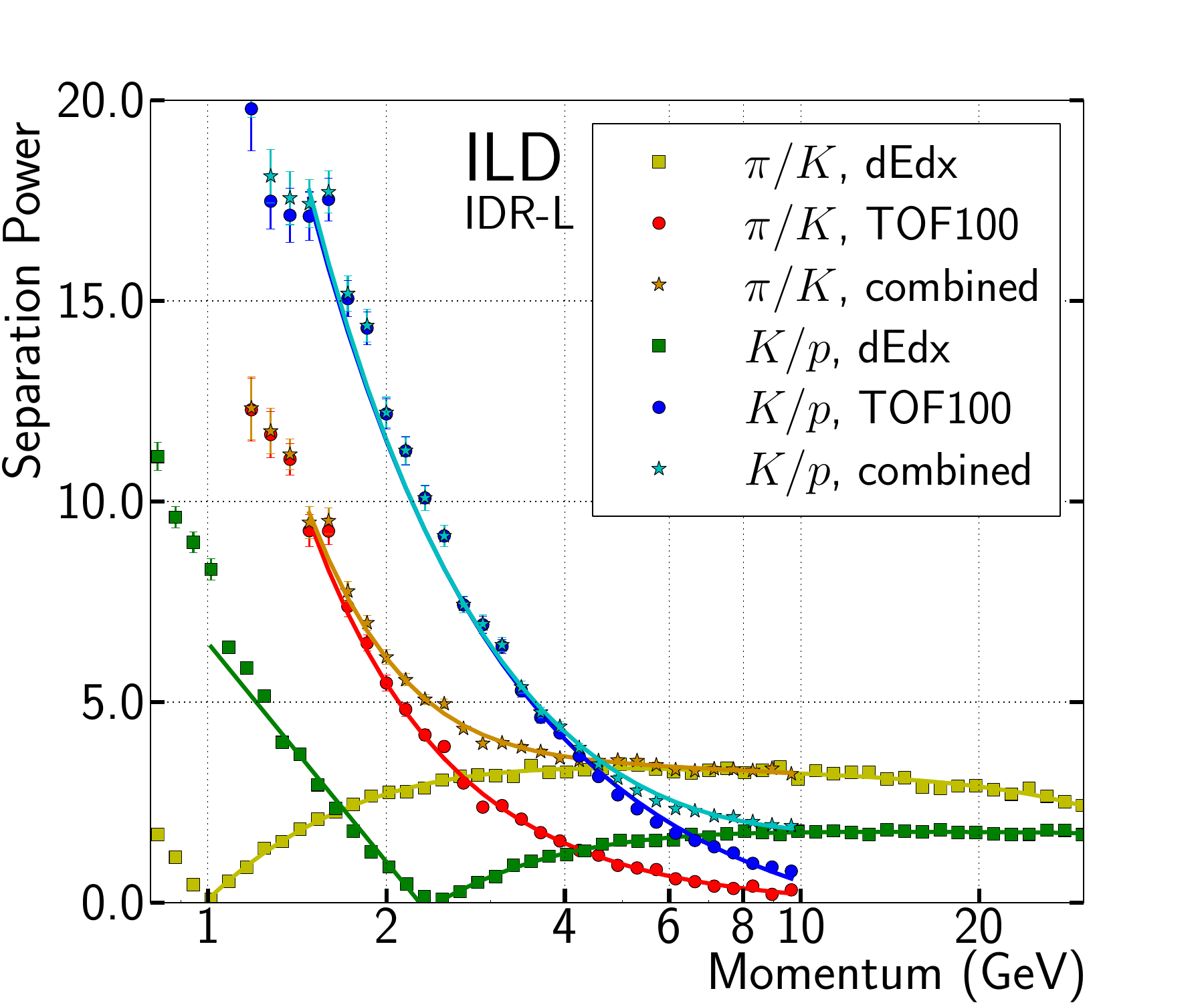}
    \end{tabular}
    \caption{ Left: Simulated resolution in $1/p_t$ as a function of the momentum for single muons. The different curves correspond to different polar angles - the open symbols are for a smaller version of ILD studied in~\cite{ILDConceptGroup:2020sfq}.
    Right: Simulated separation power (probability for a pion(kaon) to be reconstructed as a kaon(proton)) between pions and kaons, from $dE/dx$ and from timing, assuming a 100~ps timing resolution of the first ECAL layer. Figures are taken from \cite{ILDConceptGroup:2020sfq}.}
    \label{fig:momentumvsp}
\end{figure}

Inside and outside of the TPC volume a few layers of silicon detectors provide additional high resolution points, at a point resolution of around 10~$\mu \mathrm{m}$. Combined with the TPC track, this will result in an asymptotic momentum resolution of $\delta p_t / p_t^2 = 2 \times 10^{-5}$ (GeV/c)$^{-1}$ for the complete system. Since the material in the system is very low, a significantly better resolution at low momenta can be achieved than is possible with a silicon-only tracker. The achievable resolution is illustrated in Fig.~\ref{fig:momentumvsp}, where the $1/p_t$-resolution is shown as a function of the momentum of the charged particle. In the forward direction, extending the coverage down to the beampipe, a system of two inner pixel disks (point resolution 3~$\mu$m), followed by five strip disks (resolution 7~$ \mu$m) provide tracking coverage down to the beam-pipe. 

The silicon layer outside the TPC could be instrumented by sensors with excellent timing precision to measure particles' time-of-flight.
Several sensor technologies which would be applicable 
are currently under active R\&D within the DRD3 collaboration, including Trench-Isolated Low Gain Avalanche Detectors (TI-LGADs), AC-coupled LGADs (AC-LGADs), Resistive DC-coupled Silicon Detectors (RSD-LGADs), and Inverse Low Gain Avalanche Detectors (iLGADs).

In parallel, the development of low-power front-end readout electronics is essential to ensure power-efficient operation while preserving the timing resolution, a critical factor for large-scale detectors. Sensor concepts relying on signal-sharing mechanisms (such as AC-LGADs and RSD-LGADs) which enable a reduced number of readout channels may offer advantages in optimizing the balance between spatial resolution, power dissipation, and timing performance.

The TPC 
also enables the identification of particle type by the measurement of the specific energy loss, $dE/dx$, for tracks at intermediate momenta~\cite{Hauschild:2000eg}. The achievable performance is shown in Fig.~\ref{fig:momentumvsp} (right). The relatively new approach of cluster counting in the TPC promises a significantly improved resolution. If the outer 
 silicon layers can provide timing with about 100~ps resolution, time of flight measurements can provide additional information, which is particularly effective in the momentum regime which is problematic for $dE/dx$, as shown in Fig.~\ref{fig:momentumvsp} (right). 

The design and performance of the TPC has been the subject of intense {{R\&D}} over the last 15 years. Several Micro Pattern Gas Detector (MPGD) technologies for the readout of the TPC have been successfully developed, and have demonstrated the required performance in test beam experiments. The readout of the charge signals is realised either through traditional pad-based systems, or by directly attaching pixel readout-ASICs to the endplate. A large volume field cage has been built to demonstrate the low mass technology needed to meet the 10\% X$_0$ goal discussed above. Most recently the performance of the specific energy loss, $dE/dx$, has been validated in test beam data. Based on these results, the TPC technology is sufficiently mature for use in the ILD detector, and can deliver the required performance when operating at ILC (see e.g. \cite{Attie:2016yeu,Bouchez:2007pe}).

To operate at the FCC-ee, the central tracker must be re-optimized, with special attention given to running under Z-pole conditions. At luminosity of up to $2 \times 10^{36}$~$\mathrm{cm}^{-2} \mathrm{s}^{-1}$, ionization from various sources will generate a significant space charge, leading to distortions that can reach the order of 1 cm.

Studies are ongoing to quantitatively estimate how large an impact these distortions will have on the final performance of the system. Simulation studies indicate that for energies above the Z, these effects will not significantly reduce the performance. On the Z further studies are needed. 

\subsection{Calorimeter System}
A very powerful calorimeter system is essential to reach the needed performance of the detector. Particle flow, which is driving the design of ILD to a significant extent, relies on the ability to separate individual particles in a jet, both charged and neutral. This puts the imaging capabilities of the system at a premium, and pushes the calorimeter development in the direction of a system with very high granularity. A highly granular sampling calorimeter is the solution to this challenge~\cite{Sefkow:2015hna}. The conceptual and technological development of the particle flow calorimeter have been largely done by the CALICE collaboration (for a review of recent CALICE results see {e.g.} \cite{Grenier:2017ewg}). 

ILD has chosen a sampling calorimeter readout with silicon diodes as one option for the electromagnetic calorimeter. Diodes with pads of about $(5 \times 5)$ mm$^2$ are used, to sample a shower up to 30 times in the electromagnetic section. In 2018, a test beam experiment demonstrated the large scale feasibility of this technology, by showing not only that the anticipated resolution can be reached, but also by demonstrating that a sizeable system can be built and operated. The test has provided confidence that scaling this to an ILD-sized system will be possible. An alternative 
using $(50 \times 50)\mu\text{m}^2$ MAPS-based active layer is also considered \cite{Alme:2022hxo}.

As an alternative to the silicon based system, sensitive layers made from thin scintillator strips are also investigated (ScW-ECAL). Orienting the strips perpendicular to each other has the potential to realize an effective cell size of $5\times 5$~mm$^2$, with the number of read out channels reduced by an order of magnitude.A full 32-layer prototype was recently constructed by a joint effort with the CEPC-ECAL group. It was tested in beams at CERN-PS/SPS in 2022 and 2023 and its performance is being evaluated \cite{scecal1}.

For the hadronic part of the calorimeter of the ILD detector, two technologies are studied, based on either silicon 
photomultiplier
(SiPM) on scintillator tile technology~\cite{Simon:2010mi} or resistive plate chambers~\cite{Laktineh:2010zsa}. The SiPM-on-tile option has a  moderate granularity, with $3 \times 3$ cm$^2$ tiles, and provides an analogue readout of the signal in each tile (AHCAL). The RPC technology has a better granularity, of $1 \times 1$ cm$^2$, but provides only 2-bit amplitude information (SDHCAL). For both technologies, significant prototypes have been built and operated. Both follow the engineering design anticipated for the final detector, and demonstrate thus not only the performance, but also the scalability of the technology to a large detector. 

While the original design of the system was done for the ILC, its adaptation to running at the FCC-ee is under study. The main challenge in the adaptation lies in coping with the continuous readout and with the much higher rate, bandwidth, power, and cooling requirements, without compromising granularity and compactness. For a circular option, the readout rates and, more critically, heat dissipation have to be reassessed, a work on-going; the preliminary conclusions hint strongly at the need for an active cooling, especially for the ECAL. Practical cooling aspects were reviewed for the HCAL~\cite{Grondin:2228602}. More recently, studies toward a thin and uniform active cooling for the ECAL have started.

The potential of adding precise timing to the calorimeters, or at least to a few layers, is actively being investigated – from the sensor to physics performances – for most of the options, T-SDHCAL, AHCAL, SiW-ECAL, and ScW-ECAL. For the T-SDHCAL, this is accompanied by new developments in sensors, moving from RPCs to multigap RPCs, which offer better time resolution of only tens of picoseconds and enhanced rate capability. Other alternatives under R\&D, such as GEM or Micromegas, could also be promising options. For the AHCAL, the ongoing construction of a significant calorimeter in this technology for the CMS upgrade is providing important information on system design and performance, which will go a long way towards an optimized design for 
a future lepton collider detector.

A new generation of ASICs is also under development at the Omega laboratory, focusing on reducing power consumption and data volume, including auto-trigger and data-driven readout, as well as improving time resolution down to tens of picoseconds. Transitioning from the AMS130 CMOS technology used in the first ASICs to more modern technologies, such as TSMC 130 or 65, could meet the sensor requirements for all types of future lepton colliders.

\begin{figure}[th]
    \centering
    \begin{tabular}{lcr}
    \includegraphics[width=0.4\hsize]{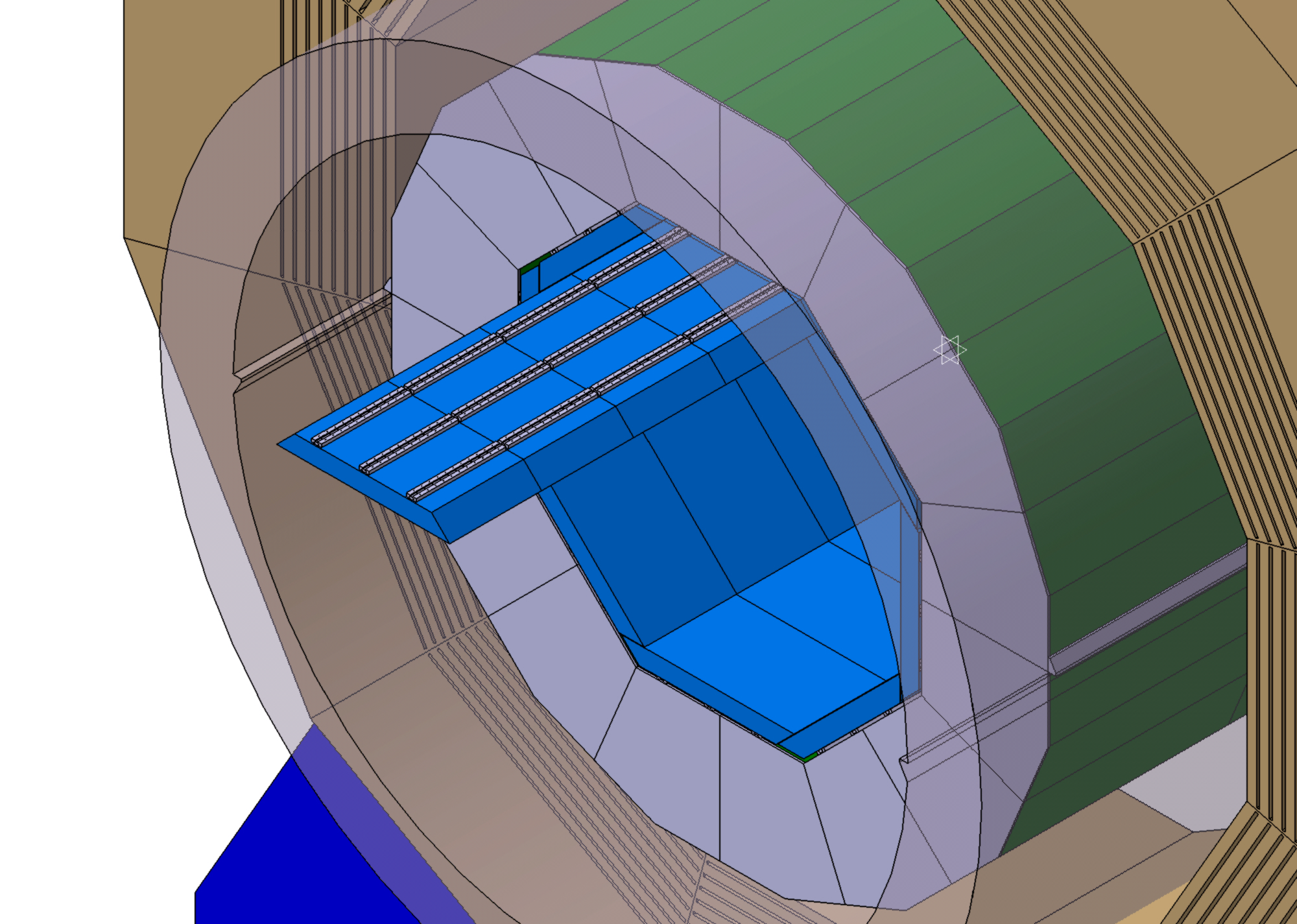} & ~~~~ &
    \includegraphics[width=0.4\hsize]{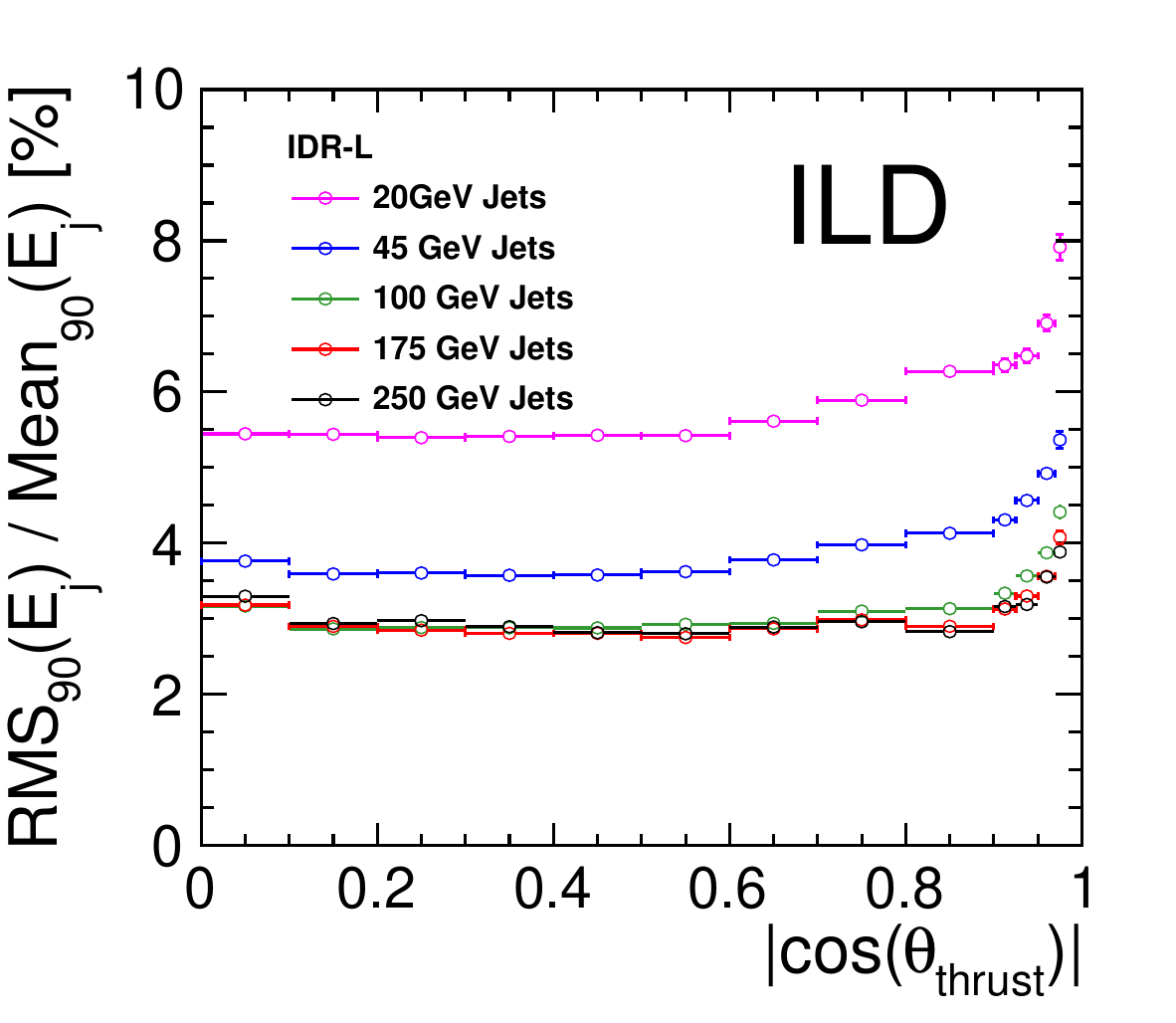}\\
    \end{tabular}
    \caption{Left: Three-dimensional rendering of the barrel calorimeter system, with one ECAL module partially extracted. Right: Particle flow performance, measured as the energy resolution in two-jet light flavour events, for different jet energies as a function of $\cos\theta)$. The resolution is defined as the rms of the distribution truncated so that 90\% of the total jet energy is contained inside the distribution. Figures are taken from \cite{ILDConceptGroup:2020sfq}.}
    \label{fig:pflow}
\end{figure}

It has been a major success in the past years that the technologies needed for a true particle flow calorimeter have been successfully demonstrated in a design which is suitable for the ILD detector. With this demonstration, a major hurdle towards the realization of ILD has been overcome~\cite{Sefkow:2018rhp}. The simulated particle flow performance is shown in Fig.~\ref{fig:pflow}.

The iron return yoke of the detector, located outside of the coil, is instrumented to act as a tail catcher and as a muon identification system. Several technologies are possible for the instrumented layers. Both RPC chambers and scintillator strips readout with SiPMs have been investigated. Up to 14 active layers, located mostly in the inner half of the iron yoke (see Table~\ref{ild:tab:barrelpara} and Fig.~\ref{fig:ILD} for more details) could be instrumented.

\subsection{The Forward System}
Three rather specific calorimeter systems are foreseen for the very forward region of the ILD detector~\cite{Abramowicz:2010bg}. LumiCal is a high precision fine sampling silicon tungsten calorimeter primarily designed to measure electrons from Bhabha scattering, and to precisely determine the integrated luminosity \cite{Smiljanic:2024twn}. The LHCAL (Luminosity Hadronic CALorimeter) just outside the LumiCal extends the reach of the endcap calorimeter system to smaller angles relative to the beam, and closes the gap between the inner edge of the ECAL endcap and the luminosity calorimeter, LumiCal.  Below the LumiCal acceptance, where background from beamstrahlung rises sharply, BeamCal, placed further downstream from the interaction point, provides added coverage and is used to provide a fast feedback on the beam position at the interaction region. As the systems move close to the beampipe, the requirements on radiation hardness and on speed become more and more challenging. Indeed this very forward region in ILD is the only region where radiation hardness of the systems is a key requirement. In placing the different detector components, particular care has been taken to allow the bulk of the beamstrahlung photons and pairs to leave the detector through the outgoing beampipe.
In particular the BeamCal has been positioned in such a way that backscattering of particles from the BeamCal face into the active part of the detector is minimised. As the inner region of the detector at the FCC-ee is very different, the layout of the forward system will need to be re-optimized. Work on this has started. 

\section{The Software Environment for ILD}

ILD has from the start actively contributed to the community driven software project iLCSoft~\cite{bib:ilcsoft} and more recently to the larger Key4hep~\cite{Key4hep:2022jnk} software ecosystem. A strong focus has always been on defining realistic simulation models of the ILD detector that are defined in DD4hep~\cite{bib:dd4hep} used to produce event samples in full simulation with Geant4. 
These event samples are used in physics analyses to obtain sensitivities to various aspects of particle physics, to develop data reconstruction algorithms, to optimise the layout of the detector or to investigate the potential advantages of new or improved technologies.

The WHIZARD~\cite{whizard:Kilian:2007gr} event generator is typically used to generate the hard scattering event, interfaced with CIRCE2 to describe the beam energy spectrum at ILC or FCC-ee, and including the effects of initial state radiation.
Pythia is then used to describe hadronisation and final state radiation, and Geant4 
to simulate the detailed detector. 
The Marlin software package is then used to process the simulated events. 
A number of ILD models have been developed in DD4hep as part of the detector optimisation activities using the chain described above.  

An alternative approach based on fast simulation using the SGV~\cite{sgv:Berggren:2012ar} is used in the case of very high statistics samples for which the CPU cost or time of full Geant4 simulation is prohibitive. 

A suite of digitisation and reconstruction algorithms has been developed within the Marlin framework. These convert the energy deposits in sensitive detectors simulated by Geant4 into realistic detector signals that are validated in comparisons to the performance of real sub-detector prototype test beam results. These signals are then reconstructed into particle tracks, calorimeter clusters and then to particle flow objects corresponding to single particles, followed by clustering into multiple-particle systems such as hadronic jets, tau leptons or converted photons.
Superimposing the effects of background processes such as beamstrahlung and photo-production of hadrons is the job of dedicated overlay processors. Algorithms for particle identification (leptons, different hadron flavours) are run, as well as
jet flavour identification. 

Over the past years, large samples of events from Standard Model processes have been produced at
various centre-of-mass energies, corresponding to proposed ILC energy points 250, 350, 500, 550, and 1000~GeV. These large-scale productions have used grid computing resources, organised via the LCDirac infrastructure. These samples have been instrumental in ensuring that the majority of ILD analyses are done with fully simulated event samples.

All tools used by ILD are incorporated in Key4hep, where work is currently ongoing to transfer some of the older tools to more modern replacements, including incorporating the seamless use of modern AI tools, further strengthening the use of Key4hep in the workflow. 

Computational 
resource estimates based on the software chain above including background simulations suggest that the total raw data rate of the ILC will be $\approx 1.5$~GB/s and the total estimated storage needs will be a few tens of PB/y~\cite{ILCESU1}, at least an order of magnitude smaller than for the LHC.


\section{Science with ILD}

ILD has been designed to operate with electron-positron collisions between 90~GeV and 1~TeV. The science goals of the future lepton collider have been recently reviewed in detail in \cite{ECFA-studyreport}, and will not be repeated here. The ILD concept group contributed many results to this report and it should be pointed out that most of them  were based on fully simulated events, using a realistic detector model and advanced reconstruction software, and in many cases included estimates of key systematic effects. 
%
The detector model used for the ILD studies was tested against performance of the prototype detectors. The key performance numbers for the vertexing, tracking and calorimeter systems are all based on results from test beam experiments. Key aspects of the particle flow performance, includes the single particle resolution for neutral and charged particles, the particle separation in jets, the matching between tracking and calorimetry, were also verified in dedicated measurements. 

A significant number of benchmark studies were undertaken to fully understand the performance of the ILD detector, to determine in particular the correlations between detector response, reconstruction results and science objectives, and to optimise the detector design parameters. 
The center-of-mass energy for the first stage of the lepton collider is about 250\,GeV, nevertheless, the ILD detector is designed to meet the more challenging requirements of higher center-of-mass energies, since major parts of the detector, e.g.\ the coil, the yoke and the main calorimeters will not be replaced when upgrading the accelerator. Therefore, benchmark analyses were also performed at a center-of-mass energy of 500\,GeV and even 1\,TeV.
The potential of new features incorporated in the detector design, e.g.\ time-of-flight measurement, was also considered. 
Results of all these studies were published in the ILD Interim Design Report~\cite{ILDConceptGroup:2020sfq}. 

\begin{figure}[th]
    \centering
    \includegraphics[width=0.4\hsize,trim=0 -1cm 0 0,clip]{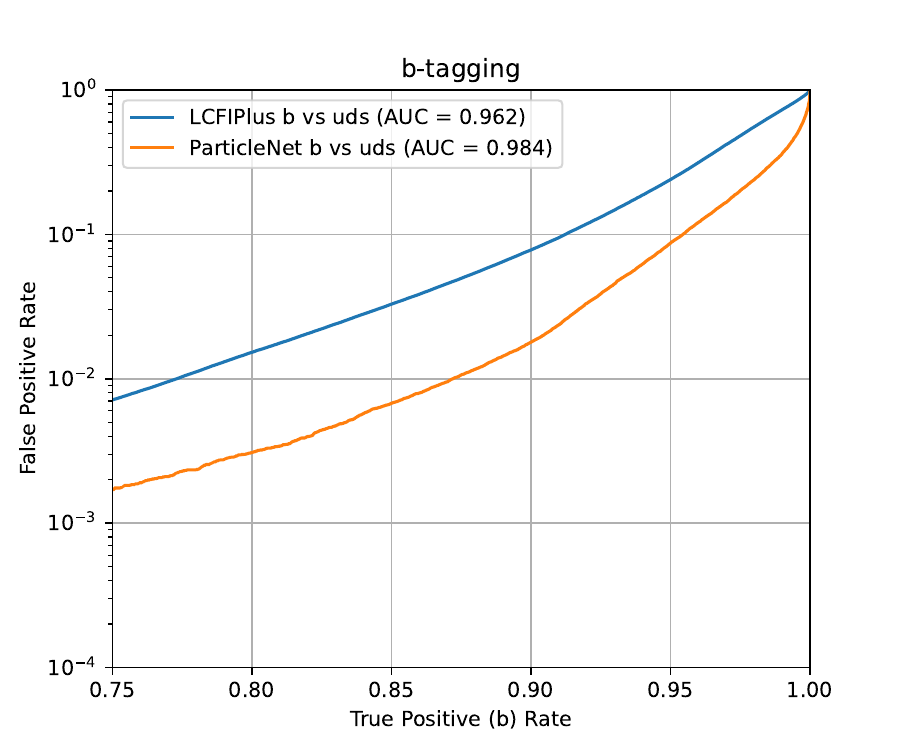} 
    \includegraphics[width=0.51\hsize]{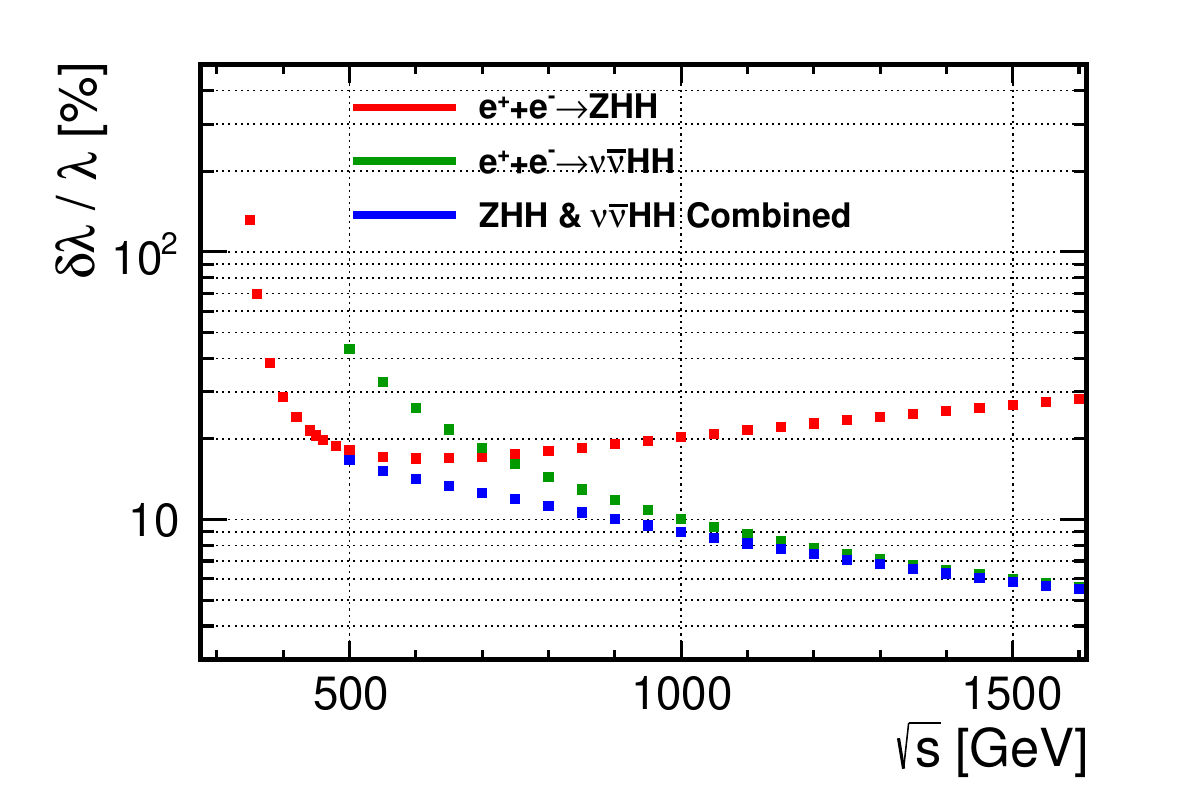}
    \caption{Left: b-tagging performance in the ZHH analysis for the LCFIPlus framework \cite{Suehara:2015ura} and new algorithm based on ParticleNet \cite{Qu:2019gqs}.
    Right: expected precision of the direct tri-linear Higgs coupling measurement at the ILC as a function of the centre-of-mass energy \cite{ECFA-studyreport}.}
    \label{fig:hself}
\end{figure}

%
The determination of the Higgs self-coupling is considered as one of the essential measurements to be done at the future lepton collider \cite{deBlas:2024bmz}. This quantity can be directly measured only at energies above 500~GeV, for example, in the Higgs pair-production process, while at lower energy, indirect determinations are possible.  
Significant progress has been obtained recently in ILD studies on the prospects for Higgs pair-production measurement, which allows for direct Higgs self-coupling determination in a model-independent way. 
Progress resulted from many improvements in the reconstruction and analysis, including improved jet clustering, new flavour tagging and particle identification algorithms,  dedicated corrections for neutrinos emitted in semi-leptonic heavy quark decays, kinematic fit and event classification based on matrix elements \cite{Radkhorrami:2021fbp,Einhaus:2022bnv,Tagami:2024gtc,Suehara:2024qoc}.
As an example, improvement in the b-tagging performance with the algorithm based on ParticleNet \cite{Qu:2019gqs}, as compared to that of LCFIPlus framework \cite{Suehara:2015ura} used previously, is shown in Fig.~\ref{fig:hself} (left).
Shown in Fig.~\ref{fig:hself} (right) is the final result of the study: expected precision of the direct tri-linear Higgs coupling measurement at ILD as a function of the centre-of-mass energy.
For a linear collider facility at CERN running at $\sqrt{s}=550$\,GeV the expected precision on the coupling measurement is 15\% (Higgs-strahlung and WW fusion channels combined) \cite{ECFA-studyreport}, compared to the earlier estimates for ILD at ILC of 26.6\% \cite{Durig:2016jrs}.

\begin{figure}[th]
    \centering
    \includegraphics[width=0.45\hsize]{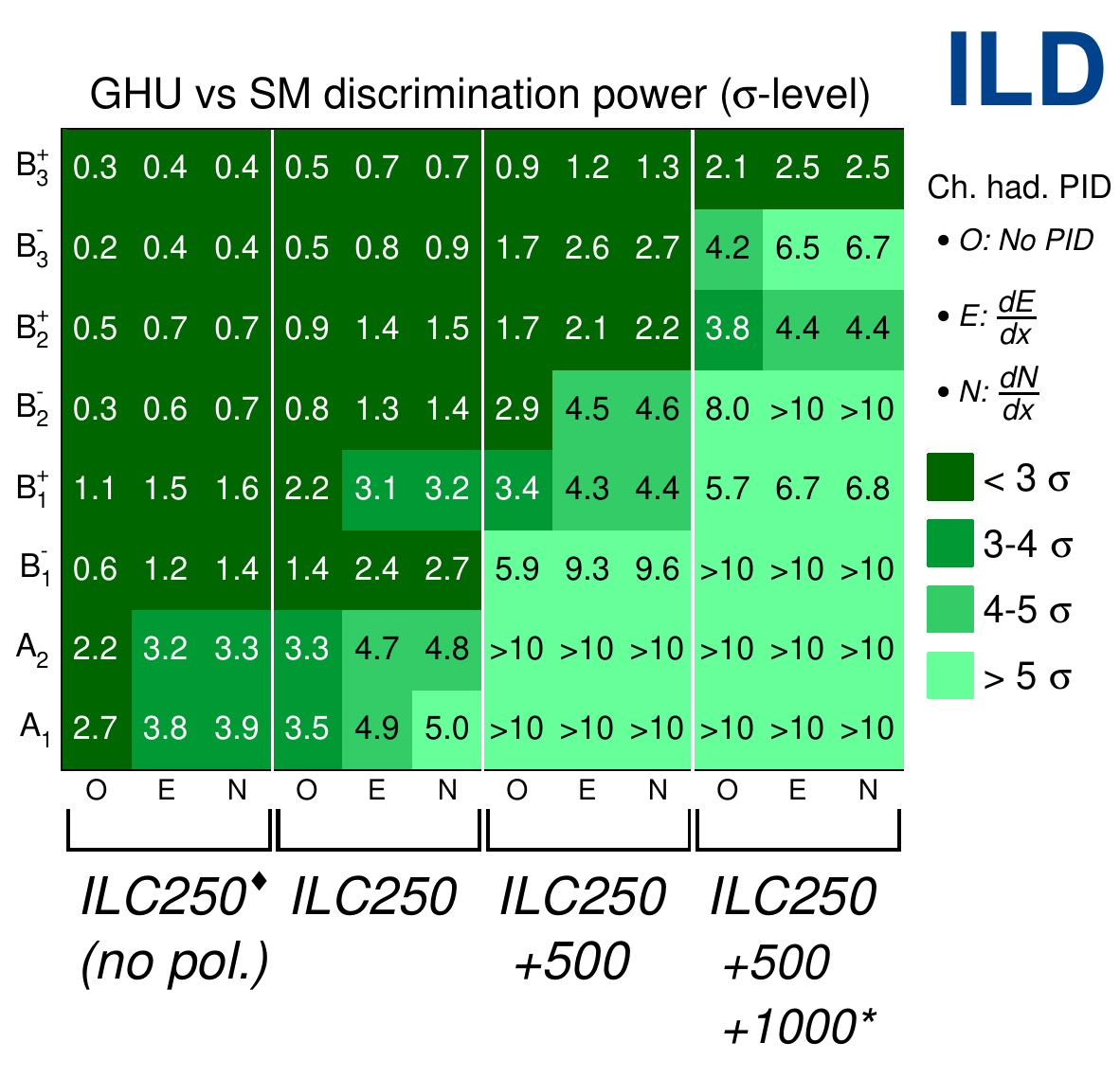} 
    \includegraphics[width=0.45\hsize]{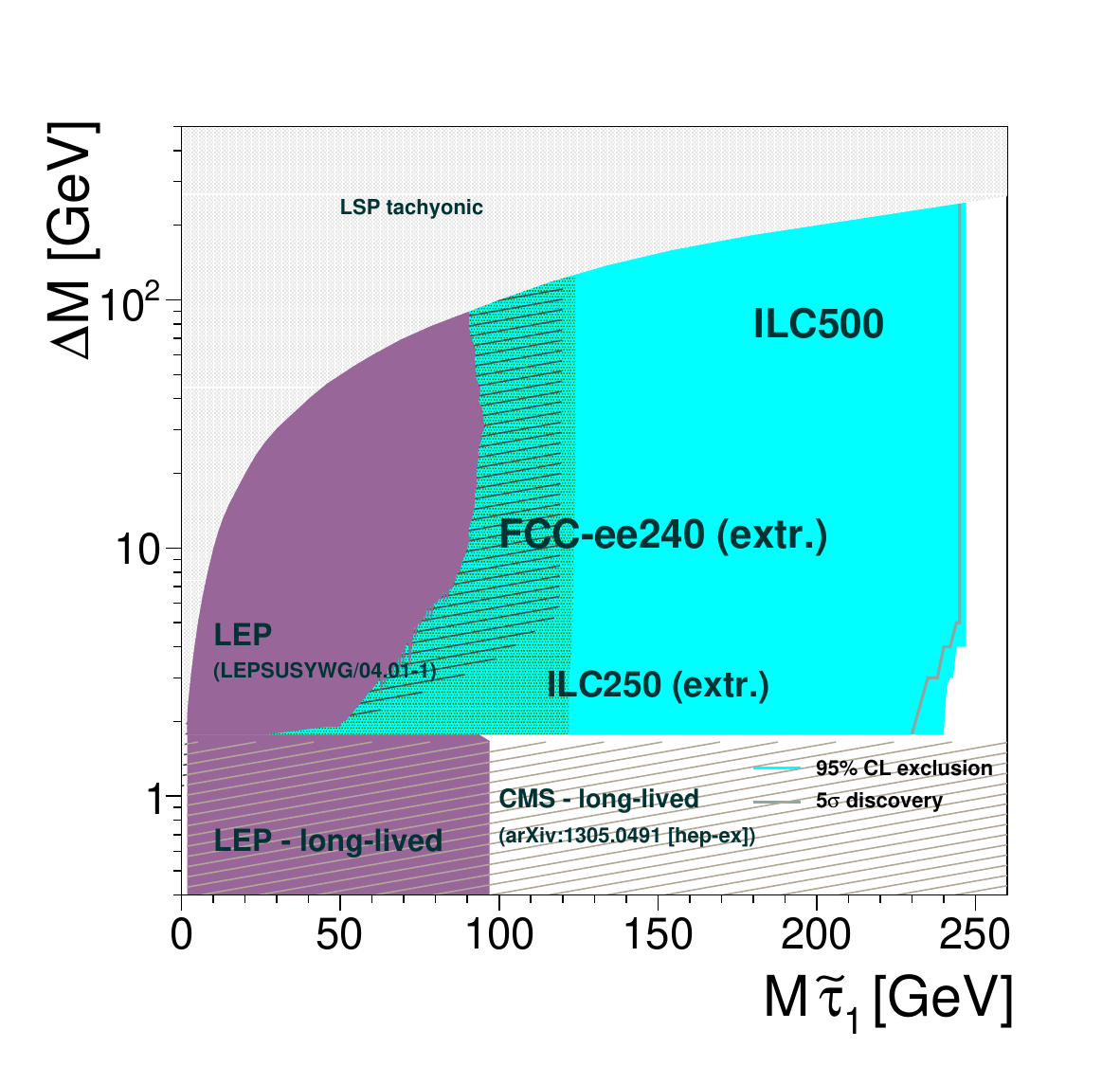}
    \caption{Left: statistical discrimination power between different gauge-Higgs unification (GHU) scenarios and the SM, for different ILC running scenarios and three different charged
    hadron particle identification capabilities (PID) considered:
    O for no PID used, $E$ for PID based on the $dE/dx$ reconstruction, 
    and $N$ for an optimized TPC design with cluster counting reconstruction $dN/dx$.
    Right: the expected discovery and exclusion reach for the $\tilde{\tau}$ NLSP search at the 500~GeV ILC.}
    \label{fig:pid_stau}
\end{figure}

Developments in the detector design and in analysis methods are also very important for precision studies and BSM searches.
Considered in the recent study \cite{Irles:2024ipg} was the impact of the TPC design on the efficiency of the charged hadron identification in the quark-antiquark forward-backward asymmetry measurement in heavy quark pair-production.
A novel, optimised TPC design, with higher readout granularity, allows for the use of the cluster counting reconstruction ($dN/dx$) method for charged-hadron identification, resulting in better resolution than the mean energy loss ($dE/dx$) measurement used in the earlier studies. 
The impact of the charged hadron particle identification (PID) on the size of the statistical discrimination power between different gauge-Higgs unification (GHU) scenarios \cite{Funatsu:2014fda,Funatsu:2019xwr} and the Standard Model is shown in Fig.~\ref{fig:pid_stau} (left). In particular in the studies of reactions involving charm quarks PID has the largest impact, and thus improvements by using $dN/dx$ are seen most strongly. 
%
%

First physics studies has also been completed comparing the ILD performance at a linear and at a circular collider. The most relevant difference  here is the design of the forward region of the detector. For a circular collider, elements from the final focus system reach further into the active volume of the detector, and limit to total hermeticity of the detector. Studies which strongly rely on the forward region of the detector are thus most strongly affected. In addition polarisation of the lepton beams is currently not anticipated for a circular collider. 
In Fig.~\ref{fig:pid_stau} (right) the expected discovery and exclusion reach for the $\tilde{\tau}$ NLSP search at the 500\,GeV ILC is shown, together with projections to ILC at 250\,GeV and FCC-ee at 240\,GeV. 
The exclusion reach of FCC-ee is significantly smaller, in the low $\Delta M$ region in particular, mainly because of the worse coverage at low angles and the lack of beam polarisation.

\section{Integration of ILD}
A detailed concept has been worked out for the ILD integration. This includes a mechanical model of the detector, and of the infrastructure needed to operate the detector, and a model of how the detector would be assembled and installed. 

The current model assumes an initial assembly of the detector on the surface, similar to the construction of CMS at the LHC. A vertical shaft from the surface into the underground experimental cavern allows ILD to be lowered in five large segments, corresponding to the five yoke rings.

ILD is self-shielding with respect to radiation and magnetic fields to enable the operation and maintenance of equipment surrounding the detector, {e.g.} cryogenics. The current design of ILD was strongly influenced by the proposal to operate two detectors at the ILD in a push-pull mode. In case this requirement is released, the integration concept of the ILD detector could be re-optimised.

The cost of the ILD detector has been estimated in a dedicated and detailed costing study in 2012. The cost has been last updated in 2019 for the ILD IDR \cite{ILDConceptGroup:2020sfq}. The total detector cost is about US\$390 million 
in 2012 costs. The cost of the detector is strongly dominated by the cost of the calorimeter system and the yoke, which together account for about $60\%$ of the total cost. 

\section{The ILD Concept Group}
The ILD concept group comprises around 275 people from 59 member institutes, and a few individual guest members, from around the world. ILD has given itself a structure, with groups who want to join do sign a memorandum of participation. Scientists who want to participate in ILD but whose institutes cannot join have the option to ask for guest membership in ILD. 

\begin{figure}
    \centering
    \includegraphics[width=0.9\hsize]{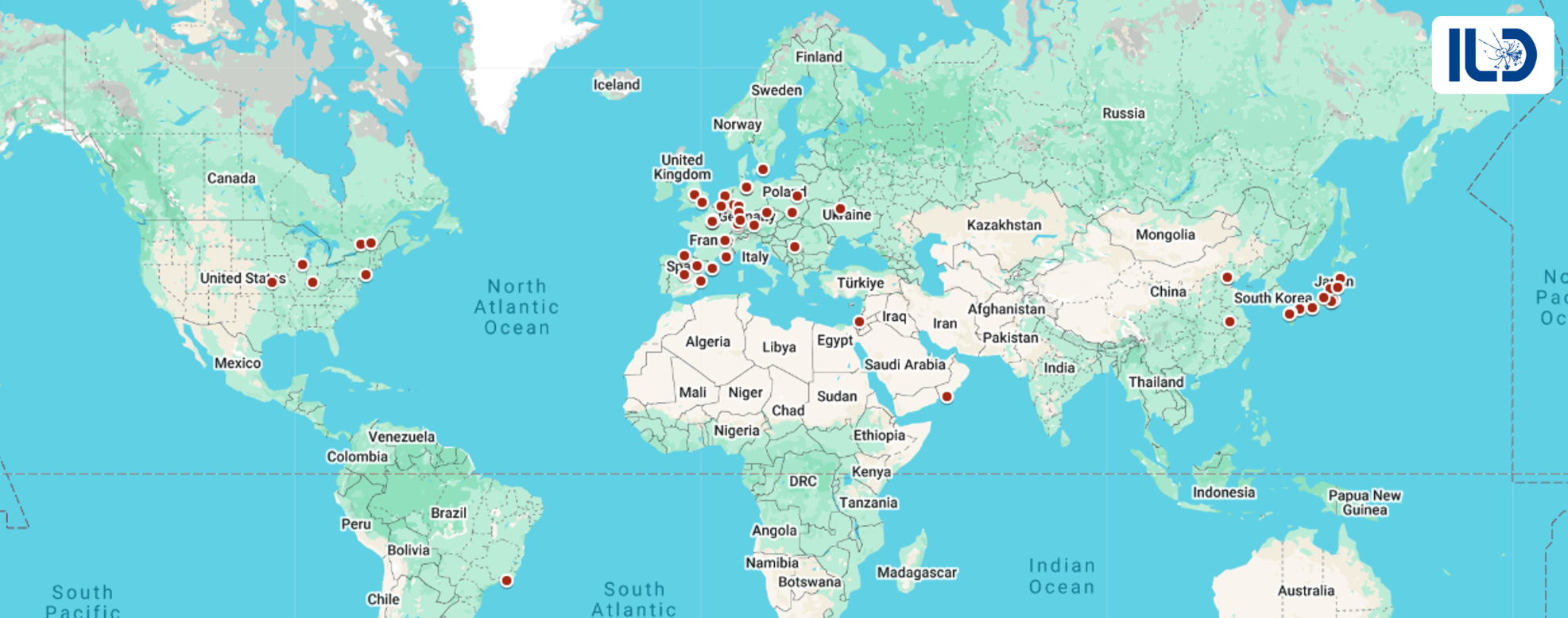}
    \caption{Locations of the ILD member institutes, as of March 2025.}
    \label{ild-fig-membermap}
\end{figure}

 A map indicating the location of the ILD member institutes is shown in Fig.~\ref{ild-fig-membermap}.

\section{Conclusion and Outlook}
The ILD 
concept is a well developed integrated detector optimised for use at a future electron-positron collider. It is based on advanced detector technology, and driven by the science requirements at future lepton collider. Most of its major components have been fully demonstrated through prototyping and test beam experiments. The physics performance of ILD has been validated using detailed simulation systems. It is the expressed intention of the ILD concept group to contribute a detector to a future lepton collider, regardless of how and where it is going to be implemented.

\section{References}
\bibliography{ILD}

\clearpage


\end{document}